\documentclass[preprint,showpacs,preprintnumbers,amsmath,amssymb]{revtex4}
\usepackage{graphicx}

\begin{document}

\title{Rectified voltage induced by a microwave field in a confined two-dimensional electron gas with a mesoscopic static vortex}
\author{D.Schmeltzer and Hsuan-Yeh Chang}
\affiliation{Department of Physics,\\City College of the City University of New York,\\
New York NY 10031}
\date{\today}

\begin{abstract}
We investigate the effect of a microwave field on a confined two dimensional electron gas which contains an insulating region comparable to the Fermi wavelength. The insulating region causes the electron wave function to vanish in that region. We describe the insulating region as a static vortex. The vortex carry a flux which is determined by vanishing of the charge density of the electronic fluid due to  the insulating region. The sign of the vorticity for a hole is opposite to the vorticity for adding additional electrons.  The  vorticity   gives rise to non-commuting kinetic momenta. The two dimensional electron gas is described as fluid with a density which obeys the Fermi-Dirac statistics. The presence of the confinement potential gives rise to vanishing kinetic momenta in the vicinity of the classical turning points. As a result the Cartesian coordinate do not commute and gives rise to a Hall current which in the presence of a modified Fermi-Surface caused by the microwave field results in a rectified voltage. Using a Bosonized formulation of the two dimensional gas in the presence of insulating regions allows us to compute the rectified current. The proposed theory may explain the experimental results recently reported by J. Zhang et al.
\end{abstract}

%\pacs{}

\maketitle

The topology of the ground state wave function plays a crucial role in determining the physical properties of a many-particle system. These properties are revealed through the quantization rules. It is known that Fermions and Bosons obey different quantization rules, while the quantized Hall conductance [1] and the value of the spin-Hall conductivity are a result of non-commuting Cartesian coordinates [2]. Similarly the phenomena of quantum pumping observed in one-dimensional electronic systems [3-5] is a result of a space-time cycle  and can be expressed in the language of non-commuting frequency $\omega=i\partial_{t}$ and coordinate $x=i\partial_{k}$ as shown in ref.[6] .

Recently, the phenomena of rectification current $I_r(V) = [ I(V) + I(-V) ] / 2$ has been proposed as a DC response to a low-frequency AC square voltage resulted from a strong $2 k_F$ scattering in a one dimensional Luttinger liquid [12].

In a recent experiment [7], a two-dimensional electron gas (2DEG) GaAs with three insulating antidots has been considered. A microwave field has been applied, and a DC voltage has been measured. The experiment has been performed with and $without$ a $magnetic$ field. The major result which occurs in the $absence$ of the $magnetic$ field is a change in sign of the rectified voltage when the microwave frequency varies from 1.46 GHz to 17.41 GHz. This behavior can be understood as being caused by the antidots, which create obstacles for the electrons.

We report in this letter a proposal for rectification.
In section $A$ we present a theory which show that rectification can be viewed as a result of non-commuting coordinates.
In section $B$ we present a qualitative model for rectification, namely the presence of vanishing wave function is described by a vortex
which induces non-commuting kinetic momenta.  The sign the vorticity is determined by the vanishing of the electronic density.  The electronic fluid can be seen as a hard core boson which carry flux, the removal of charge  caused by the insulating region is equivalent to a decrease of flux with respect  the flux of the uniform fluid.   Including in addition a confining potential we obtain regions where the momentum vanishes. The combined effect non-commuting kinetic momenta and confinement gives rise to non-commuting cartesian coordinates.
In section $C$ we use the Bosonization method to construct a quantitative theory which gives rise to  a set of equations of motion. Constructing an  iterative solution of this equations reveals the phenomena of rectifications explained in sections $A$ and $B$.

\vspace{0.2 in}

\texttt{A- Rectifications due to non-commuting coordinates}

\vspace{0.2 in}

Due to the existence of the obstacles, the wave function of the electron vanishes in the domain of the obstacles. This will give rise to a change in the wave function , $ | \psi >\rightarrow |\Phi >=U^{\dagger}(\vec{K})| \psi > $ where $U^{\dagger}(\vec{K})$ is the unitary transformation (induced by the obstacle) and the coordinate coordinate representation becomes, $\vec{r} = i \frac{\partial}{\partial K} \rightarrow \vec{r} = i \frac{\vec{\partial}}{\partial K} + U^{\dagger}(\vec{K}) \frac{i \vec{\partial}}{\partial K}U(\vec{K})$ [1,2,15].
An interesting situation occurs when the wave function $|\Phi>$ has zero's [1,2,15] or points of degeneracy [16] in the momentum space .This gives rise to non -commuting coordinates [1,2,16]. As a result we will have a situation where the the commutator $[r_1,r_2 ]$ of the coordinates  is non zero.
\begin{equation}
[r_1 (\vec{K}), r_2(\vec{K})] d K^1 d K^2 = i\Omega(\vec{K}) d K^1 d K^2
\label{curvature}
\end{equation}
Using the one particle hamiltonian $h=E(\vec{K},\vec{r})$ in the presence of an external electric field  with the commutators $[r_i(\vec{K}),K_{j}]=i\delta_{i,j}$ ,$[r_1 (\vec{K}), r_2(\vec{K})]=i\Omega(\vec{K})$  one obtains [2]  the Heisenberg equations of motions ,

\begin{eqnarray}
\frac{d r_1}{d t} &=&\frac{1}{\hbar} \frac{\partial E(\vec{K}, \vec{r})}{d K^1} + \Omega(\vec{K}) \frac{d K_2}{d t} \\ \nonumber
\frac{d K_2}{d t}& = &- \frac{e}{\hbar} E_2 (t)
\label{Motion}
\end{eqnarray}
 
This equations are identical with the one obtained in ref.[15 ] where   $E(\vec{K}, \vec{r})$ is the single particle energy being in the semi-classical approximation and 
 $E_2(t)$ the external electric field.
As a result of the external electric field $E_2(t)$
 changes the velocity changes according to eq.$2$. Using the interaction picture we find,
\begin{equation}
\dot{r}_1 = \frac{1}{i\hbar}[r_{1}(t),-er_{2}(t)E_2(t)]=\frac{e}{\hbar} \Omega(\vec{K})E_2(t)\approx\frac{e}{\hbar}\Omega(\vec{K})V_2(t) / L
\label{external}
\end{equation}

 $V_2(t)$ is the voltage caused by the external field $E_2(t)$. The Fermi Dirac occupation function $\rho(\vec{K}, \vec{r}) = f_{F.D.} [ E(\vec{K}, \vec{r}) - e V_2(t) - E_F]$ in the presence of the electric field is used to sum over all the single particle states . We obtain the current density $J_1 (r)$ in the $i=1$,
\begin{equation}
J_1 (r) \simeq e \int \frac{d^2 K}{(2 \pi)^2} \dot{r}_1 (\vec{K}) \rho(\vec{K}, \vec{r}) \simeq \frac{e}{\hbar} \int \frac{d^2 K}{(2 \pi)^2} \Omega(\vec{K}) \rho(\vec{K}, \vec{r}) \frac{V_2(t)}{L}
\label{current}
\end{equation}
The result obtained in  the last equation follows directly from the non-commuting coordinates given by $\Omega(\vec{K})$ $\neq 0$. The current in eq. $1$ depends on $\rho(\vec{K}, \vec{r}) = f_{F.D.} [ E(\vec{K}, \vec{r}) - e V_2(t) - E_F]$, the Fermi- Dirac occupation function in the presence of the external voltage $ V_2(t) \simeq E_2 (t) L$ . We expand the non equilibrium density $\rho(\vec{K}, \vec{r})$ to first order in $V_2(t)$ we obtain the final form of the rectified current. $\Omega(\vec{K})$ has dimensions of a frequency and can be replaced with the help of the Larmor's theorem, by an $effective$ magnetic field $\Omega(\vec{K}) = \frac{e}{2 m c} B_{eff}(\vec{K})$. This allows us to replace eq. 4 by the formula.
\[
I_{1} = \frac{e^2 V_2^2}{2 \hbar m c} \int \frac{d^2 K}{(2 \pi)^2} \int \frac{d r_1}{L} B_{eff}(\vec{K}) \delta(E(\vec{K}, \vec{r}) - E_F) \propto {(V_2(t))^2} B_{eff}.
\]
\vspace{0.2 in}

\texttt{B-A model for non-commuting coordinates}

\vspace{0.2 in}

 We consider a two dimensional electron gas (2DEG) in the presence of a parabolic confining potential $V_{c}(\vec{r})$.
 The 2DEG contains an insulting region of radius $D$ caused by an infinite potential $U_I(r)$ ( in the experiment the insulating region this is caused by three antidots) see figure $1a$.The effect of the insulating region of radius $D$ causes the electronic wave function $ | \tilde{\psi}(r;R) >$ to vanish for $|\vec{r} - \vec{R} | \le D$.
The spin of the electrons seems not to play any significant role,therefore we approximate the 2DEG by a spinless  charge system. Such a charged electronic system   is equivalent to a hard core  charged Boson . For Bosonic  wave function  has zero's  which  can be described as a vortex  centered at $\vec{r} = \vec{R}$ .

\begin{figure}
% G1.eps: 300dpi, width=1.96cm, height=1.21cm, bb=91 3 322 146
\includegraphics[width=3.5in]{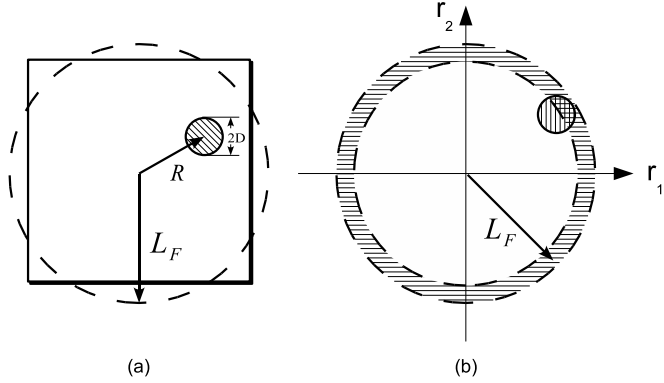}
\caption{(a) Figure 1a; (b) Figure 1b.}
\label{fig:1}
\end{figure}

We will show that the following properties are $essential$ in order to have non-commuting coordinates.

\vspace{0.1 in} 
 
    1-The $vanishing$ of the wave function for $|\vec{r} - \vec{R} | \le D$ is described by a vortex localized at $\vec{R}$.
    
\vspace{0.1 in}  
   
     2-The many particles will be described in term of a continuous Lagrange formulation [14]  $\vec{r} (\vec{u}, t)$. Here, $\vec{r} (\vec{u}, t)$ is the continuous form of $\vec{r}_\alpha (t)$, where ``$\alpha$ '' denotes the particular particle, $\alpha = 1, 2, \ldots, N$ with a density function $\rho_0(\vec{r})$, which satisfies $ N/L^2 = \int \rho_0 (\vec{u}) d^2u$ in two dimensions ($L^2$ is the two dimensional area). The coordinate $\vec{r}(\vec{u})$ and the momentum $\vec{P}(\vec{u})$ obey canonical commutation rules, $ [ r_i (\vec{u}), P_j (\vec{u}')] = i \delta_{ij} \delta^2(\vec{u} - \vec{u}')$.
     
\vspace{0.1 in}  
   
3- The parabolic confining parabolic potential
$V_c(\vec{r}) = \frac{m \omega_0^2}{2} \vec{r}^2$  provides the confining length $L_{F}$  ,see figures $1a$ and $1b$.

\vspace{0.1 in}

\vspace{0.1 in}

Using the conditions 1-3 we will show that the non-commuting coordinates emerge:

 \vspace{0.1 in}
    
 -The $vanishing$ of the $wave$ $function$.
 
 In the literature it was established that the vanishing of the Bosonic wave function gives rise to a multivalued phase and vorticity . See in particular the derivation given in ref. [9]. The $vortex$ (the insulating region ) gives rise to $non$-$commuting$ $kinetic$ $momenta$ , $[\Pi_{1}(\vec{r}),\Pi_{1}(\vec{r'})]\neq 0$ where , $\vec{\Pi} = \vec{K} - \vec{\partial} \theta(\vec{r}; \vec{R})$ and the phase $\vec{\partial} \theta(\vec{r}; \vec{R})$ is caused by the localized vortex [8,9,10].This result is obtained in the following way:
 
 In the presence of a vortex
the single particle operator is parametrize as follows:
$
\tilde{\psi}(\vec{r}; \vec{R})= \frac{|\vec{r} - \vec{R}|}{D} e^{i \theta(\vec{r}, \vec{R})} \psi(\vec{r})$
for $|\vec{r} - \vec{R}| < D$, and $\tilde{\psi}(\vec{r}; \vec{R})= e^{i \theta(\vec{r}, \vec{R})} \psi(\vec{r})
$
for $|\vec{r} - \vec{R}| > D$. The field $\psi(\vec{r})$ is a regular hard core boson field and $\theta(\vec{r};\vec{R})$ is a multivalued phase. As a result, the Hamiltonian $\hat{h}_0 = \frac{\hbar^2}{2 m} \vec{K}^2 + U_I(\vec{r})$ and the field $\tilde{\psi}(\vec{r}; \vec{R})$ are replaced by the transformed Hamiltonian:
\begin{equation}
h_0 =\frac{\hbar^2}{2 m} (\vec{K}- \vec{\partial} \theta(\vec{r};\vec{R}) )^2
\label{hamil}
\end{equation}

 The momentum $\vec{K}$ is replaced by the kinetic momentum , $\vec{\Pi} = \vec{K} - \vec{\partial} \theta(\vec{r}; \vec{R})$. The derivative of the multivalued phase $\theta(\vec{r}; \vec{R})$ determines the vector potential $ \vec{A} (\vec{r}; \vec{R})=\vec{\partial} \theta(\vec{r}; \vec{R})$.
  $[\Pi_1, \Pi_2] \neq 0$.
\begin{equation}
[ \Pi_1(\vec{r}), \Pi_2(\vec{r'}) ] = i \bar{B}(\vec{r}) \delta (\vec{r}-\vec{r}')\approx i\frac{E(\vec{r},\vec{R})}{D^{2}} \delta (\vec{r}-\vec{r}')
\label{commutator}
\end{equation}
where $\bar{B}$ is an effective magnetic field due to the insulation region, which  is defined as $\bar{B}(\vec{r}) = \nabla \times \vec{A} (\vec{r}; \vec{R})$. 
\textbf{The sign of the magnetic field  $\bar{B}(\vec{r})$ is determined  by the vorticity. Following  the theory presented in ref.8  (see pages 94-99 and 222-227)  $\bar{B}(\vec{r})$ has  positive vorticity since the electronic  density  vanishes for the region   $|\vec{r} - \vec{R}| > D$ creating a $hole $ on background   density   ( see figure 13.1 page 227 in ref.8).}
 
For the remaining part of this paper we will replace the delta function by a step function $E(\vec{r},\vec{R})$ which takes the value of one for $|\vec{r}-\vec{R}|<D$ and zero otherwise.

\vspace{0.1 in}

-$The$ $many$ $particle$ $representation$.

In the presence of hard core Bosons (spinless Fermions) the momenta is replaced by $\hbar \vec{K}\rightarrow \vec{P}(\vec{u})$.
The static vortex describes the insulating region and modifies the momentum operator ,

$\vec{\Pi}(\vec{u})=\vec{P}(\vec{u})-\vec{\partial}_{u}\vartheta(\vec{r}(\vec{u});\vec{R})$

Making use of this continuous formulation, we find a similar result as we have for the single particles [9].
\begin{equation}
 [\Pi_1 (\vec{u}), \Pi_2 (\vec{u}') ] = \frac{i E(\vec{u},\vec{R})}{D^2} \delta^2 (\vec{u} - \vec{u}')
\label{commutators}
\end{equation}
Where $E(\vec{u},\vec{R})=1$ for $|\vec{u}-\vec{R}|<D$ and zero otherwise .

\vspace{0.1 in}

-$The$ $confining$ $potential$.

The   last ingredient of our theory is provided by the  confining potential and the $Fermi$ $energy$. Due to the confining potential the $kinetic$ $momentum$ has to vanishes for particles which have the coordinate close to the classical turning point  (see figure $1b$) $|\vec{r}|\approx L_{F}$ , $E_{fermi}=V_c(|\vec{r}|\approx L_{F})$ .
This lead to the following constraint problem for the kinetic momentum,
\begin{equation}
\Pi_1 (|\vec{u}|\approx L_{F})|\psi>=0
\label{constraints1}
\end{equation}

and

\begin{equation}
\Pi_2 (|\vec{u}|\approx L_{F})|\psi>=0
\label{constraints2}
\end{equation}

The kinetic momentum $\Pi_1 (|\vec{u}|\approx L_{F})$ and $\Pi_2 (|\vec{u}|\approx L_{F})$ form a $second$ $class$ $constraints$
(according to Dirac's definition [11] the commutator of the constraints has to be non-zero ) $[\Pi_1 (|\vec{u}|\approx L_{F}),\Pi_2 (|\vec{u}|\approx L_{F})]\neq 0$ if the region
$|\vec{r}|\approx L_{F}$ $overlaps$ with the vortex region $|\vec{r} - \vec{R} | \le D$. For $|\vec{u} - \vec{R} | \le D$ the $commutator$ of the $kinetic$ $momenta$ is given by $C\equiv\frac{1}{D^2}$  given by eq.7.

We define  the  matrix  $[C_{1,2}(\vec{u},\vec{u'})]^{-1}\equiv [\Pi_1 (\vec{u}), \Pi_2 (\vec{u}') ]$. Using the function $E(\vec{u},\vec{R})$ (which replaces the delta function ) and the eqs.8,9  we obtain 

\begin{equation}
[C_{1,2}(\vec{u},\vec{u'})]^{-1}\approx C^{-1}E( \vec{u}, \vec{R} )E( \vec{u'}, \vec{R} ) \delta (|\vec{r}(\vec{u})| -L_{F})\delta (|\vec{r}(\vec{u'})| -L_{F})
\label{inverse}
\end{equation}

The overlapping conditions are given by the conditions:
$E (\vec{u},\vec{R})$ is equal to one for $| \vec{u} - \vec{R} |<D$ and zero otherwise and $\delta (|\vec{r}(\vec{u}) - L_{F}) |$ describes the condition of the classical turning points.  Using eq.10 we find according to Dirac's second class constraints  [11] the following  new commutator   $[,]_{\mathcal{D}}$,

\begin{eqnarray}
[r_1(\vec{u}), r_2(\vec{u}') ]_{\mathcal{D}} &=& [ r_1 (\vec{u}), r_2(\vec{u}') ] - \int\,du'''\int\,du''[ r_1 (\vec{u}), \Pi_1(\vec{u}'') ] \left( C_{1,2} (\vec{u}'', \vec{u}''') \right)^{-1} [ \Pi_2 (\vec{u}'''), r_2(\vec{u}') ] \nonumber \\  
& \simeq & i\left[ D^2 E (\vec{u} ,\vec{R} )E (\vec{u'} ,\vec{R} )\delta (|\vec{r}(\vec{u})| -L_{F})\delta (|\vec{r}(\vec{u'})| -L_{F})\right] \delta ( \vec{r}'(\vec{u}) - \vec{r}(\vec{u}') )  
\end{eqnarray}
%\equiv  i\Omega(\vec{u};\vec{R}) \delta (\vec{r}'(\vec{u}) - \vec{r}(\vec{u}'))  

For  $|\vec{r}-\vec{R}|<D$   we define a field $\Omega(\vec{u};\vec{R})$  trough the equation,

$[ D^2 E (\vec{u} ,\vec{R} )E (\vec{u'} ,\vec{R} )\delta (|\vec{r}(\vec{u})| -L_{F})\delta (|\vec{r}(\vec{u'})| -L_{F})]
\equiv  \Omega(\vec{u};\vec{R}) $

This means that $\Omega(\vec{u};\vec{R})$ is approximated by  $\Omega\propto D^2$ for $|\vec{r}-\vec{R}|<D$ .

\textbf{Eq. 11 shows that  the presence of the   momentum , $\vec{\Pi}(\vec{u})=\vec{P}(\vec{u})-\vec{\partial}_{u}\vartheta(\vec{r}(\vec{u});\vec{R})$
with  the  constraints given by  eqs. 8,9  gives rise to non-commuting coordinates $[r_1(\vec{u}), r_2(\vec{u}') ]_{\mathcal{D}}\neq 0$.}

Once we have the result that the coordinate do not commute we can use the analysis given in eq $4$ (and the result for the current $I_{1}$ derived with the help of equation $4$) to compute 
 the rectified current, $I_{1} \propto \frac{ e^2}{\hbar} \frac{D^2}{L^{2}_{F}} V^{2}_{2}(t)$

 This  results can be derived  in directly   using a modified Bosonization method with  a non -commuting Kack Moody algebra [16,17].

\vspace{0.2 in} 

\textbf{C- The continuous formulation for the 2DEG-A Bosonization approach}

\vspace{0.2 in}

\textbf{C1-Bosonization  for the 2DEG}
 
\vspace{0.2 in}

We introduce a continuous formulation for the 2DEG many particles system. We replace the single particle Hamiltonian $h_{total}(\vec{\Pi}, \vec{r})$ by a many electron formulation [14]. We introduce a continuous representation, namely $\vec{r} (\vec{u}, t)$. Here, $\vec{r} (\vec{u}, t)$ is the continuous form of $\vec{r}_\alpha (t)$.
The coordinate and the momentum obey $\vec{r}(\vec{u}, t=0) = \vec{u}$ and $\vec{P}(\vec{u}, t=0) = \vec{K}$
 The equilibrium Fermi-Dirac density is given by $ \rho_0 (\vec{u}) = \int \frac{d^2 K}{(2 \pi)^2} f_{F.D.} [ \frac{\hbar^2}{2 m} \vec{K}^2 + V_c(\vec{u}) - E_F] $

One of the useful description for many electrons in two dimensions  is the Bosonization method. We will modify this method [16,17] in order to introduce  the effect of the $vortex$ field  and the $confining$ $potential$ $V_{c}(\vec{u})$.

\vspace{0.2 in}

\textbf{C2-The Bosonization method in the absence of the insulating region and confining potential}

\vspace{0.2 in}

In  this section we will present the $known$ [16,17]  results for a two dimensional interacting metal  in the $absence$ of the $vortex$ field  and $confining$ $potential$   $V_{c}(\vec{u})$.
Our starting point is the $Bosonized$ form of the
2DEG given in ref. [16,17].

\vspace{0.2 in}

 $H_{F.S.}=\frac{1}{2}\int d^{2}r\int d^{2}r^{\prime
}\oint \frac{\left \vert \overrightarrow{k}_{F}^{0}\left( s\right)
\right
\vert }{\left( 2\pi \right) ^{2}}ds\oint \frac{\left \vert 
\overrightarrow{k}_{F}^{0}\left( s^{\prime }\right) \right
\vert }{\left(
2\pi \right) ^{2}}ds^{\prime }\Gamma \left( s,s^{\prime };\overrightarrow{r}-%
\overrightarrow{r}^{\prime }\right) :\delta k_{||}\left( s,\overrightarrow{r}%
\right) \delta k_{||}\left( s^{\prime },\overrightarrow{r}^{\prime }\right) :$

\vspace{0.2 in}

 Where $\Gamma \left( s,s^{\prime };\overrightarrow{r}-\overrightarrow{r}%
^{\prime }\right) $ is the Landau function for the two body interaction [17] and the notation $: :$  represents the normal order with respect the Fermi Surface. [17]. 
According to 
ref.[16,17] the F.S. is described by, $\overrightarrow{k}_{F}\left( s,%
\overrightarrow{r}\right) =\overrightarrow{k}_{F}^{0}\left( s\right) +\delta 
\overrightarrow{k}_{F}\left( s,\overrightarrow{r}\right) $ .The "normal"
deformation to the F.S. is given by, $\delta k_{||}\left( s,\overrightarrow{r%
}\right) \equiv \widehat{n}\left( s\right) \cdot \delta \overrightarrow{k}%
_{F}\left( s,\overrightarrow{r}\right) $. $"s"$ is the polar angle on the
F.S. $\overrightarrow{k}^{0}\left( s\right)$, and $\widehat{n}\left( s\right) 
$ is the normal to the F.S..
 The $commutation$ relations for the F.S. are,

$\left[
\delta k_{||}\left( s,\overrightarrow{r}\right),\delta k_{||}\left(
s^{\prime },\overrightarrow{r}^{\prime }\right) \right] =i\left( 2\pi
\right) ^{2}\widehat{n}\left( s\right) \cdot \nabla \delta ^{2}\left( 
\widehat{n}\left( s\right) \cdot \overrightarrow{r}-\widehat{n}\left(
s^{\prime }\right) \cdot\overrightarrow{r'}\right) $ $\delta \left( 
\overrightarrow{k}_{F}^{\prime 0}\left( s\right) -\overrightarrow{k}%
_{F}^{0}\left( s^{\prime }\right) \right) $ 

\vspace{0.2 in}

\textbf{C3- The modification of the Bosonization  method in the presence of a confining potential}

\vspace{0.2 in}

Following ref.[18] ( see  the last term of  eq.10 in ref. 18 ) we  incorporate into the Bosonic hamiltonian the effect of  the confining  and  external potentials. 
We parametrize the Fermi surface in terms of the polar angle $s=[0-2\pi]$ and the coordinate $\vec{u}$ .The Fermi surface momentum $K^{0}_{F}(s,\vec{u})$ given by the solution , $K^{0}_{F}(\vec{u})=\frac{2m}{\hbar^{2}}\sqrt{E_{F}-\frac{m\omega^{2}_{0}}{2}(\vec{u})^{2}}$. As a result the
the $FERMI$ $SURFACE$ ( F.S.) excitations  is given by , $\overrightarrow{K}_{F}\left( s,%
\overrightarrow{u}\right) =\vec{K}_{F}^{0}( s,\vec{u}) +\delta
\overrightarrow{k}_{F}\left( s,\overrightarrow{u}\right) $ .The "normal"
deformation to the F.S. is given by, $\delta k_{||}\left( s,\overrightarrow{u}\right) \equiv \widehat{n}\left( s , \vec{u} \right) \cdot \delta \overrightarrow{k}%
_{F}\left( s,\overrightarrow{u}\right) $ and $\widehat{n}(s,\vec{u})$ is the normal to the F.S. as a function of  the polar angle $s$ and  real space coordinate $\vec{u}$.

\textbf{Following refs.[16,17,18] 
we obtain the Bosonized hamiltonian for the many particles system in the presence of the potentials , $V_{c}(\vec{u})$ and  time dependent external potential $U^{ext}(\vec{u},t)$ .} 
 
\begin{equation}
H = \int d^{2}u\oint \frac{ |\overrightarrow{K}_{F}^{0}( s,\vec{u})|}
{( 2\pi ) ^{2}}  [\frac{\hbar |\vec{K}_{F}^{0}( s,\vec{u})|}{2 m}
(\delta k_{||}( s,\vec{u}))^2  +V_c(\vec{u})\delta k_{||}( s,\vec{u}) + (-e)U^{ext}(\vec{u},t)\delta k_{||}( s,\vec{u})]\,ds
\label{hamilbos}
\end{equation}
The new part of in the Bosonic hamiltonian is the presence of the  confining  $V_c(\vec{u}) = \frac{m \omega_0^2}{2} \vec{u}^2$  and external potential $U^{ext}(\vec{u},t)$.
$U^{ext}(\vec{u},t)$ is the external microwave radiation field,

 $E_{2}(t)=-\partial_{2}U^{ext}(\vec{u},t)=E_{c}cos(\omega t +\alpha(t)$ and
$E_{1}(t)=-\partial_{1}U^{ext}(\vec{u},t)=0$

 The $commutation$ relations for the $FERMI$ $SURFACE$ in  are given by the Kack Moody commutation relation [16,17 ]
\begin{equation}
\left[
\delta k_{||}\left( s,\overrightarrow{u}\right),\delta k_{||}\left(
s^{\prime },\overrightarrow{u'}^{\prime }\right) \right] =i\left( 2\pi
\right) ^{2}\widehat{n}(s,\vec{u}) \cdot \nabla \delta ^{2}\left(
\widehat{n}(s,\vec{u})\cdot\vec{u}-\widehat{n}(
s'\vec{u'})\cdot\vec{u'}\right) \delta \left(
\overrightarrow{K}_{F}^{ 0}( s,\vec{u}) -\overrightarrow{K}%
_{F}^{0}(s',\vec{u'})\right)
\label{kack}
\end{equation}

\vspace{0.2 in}

\textbf{C4-The Bosonization method in the presence of the insulating region  and confining potential}

\vspace{0.2 in}

This problem can be investigated using the hamiltonian given in eq.12 supplemented  by the constraints conditions imposed by the vanishing density.
 described by a vortex. Using the results given in  eq.11 one  modifies the commutation relations. This modification can be viewed as Dirac's $bracket$ due to second class constraints [11 ]. The commutator $[,]$ is replaced by Dirac 's commutators $[,]_{\mathcal{D}}$.
  The region of vanishing density  is described by the function $E\left( \overrightarrow{u%
}\right) =1$ for $|\vec{u}-\vec{R}|<D$ and zero otherwise . Using   $[r_1(\vec{u}), r_2(\vec{u}') ]_{\mathcal{D}}\neq 0$ we find that  the Dirac
commutator $\left[ \delta k_{||}\left( s,\overrightarrow{u}\right) ,\delta k_{||}\left(
s^{\prime },\overrightarrow{u}^{\prime }\right) \right]_{\mathcal{D}}$    replaces  the  commutator given in equation 13

\begin{eqnarray}
\left[ \delta k_{||}\left( s,\overrightarrow{u}\right) ,\delta k_{||}\left(
s^{\prime },\overrightarrow{u}^{\prime }\right) \right] _{\mathcal{D}} =\left[
\delta k_{||}\left( s,\overrightarrow{u}^{\prime }\right) ,\delta
k_{||}\left( s^{\prime },\overrightarrow{u}^{\prime }\right) \right]
- \nonumber \\\int d^{2}z\int d^{2}z^{\prime }  
\left[ \delta k_{||}\left( s,\overrightarrow{u}\right) ,r_{1}( 
\overrightarrow{z}) \right] ([r_{1}(\overrightarrow{z}), r_{2}( \overrightarrow{z}
^{\prime }) ]_{\mathcal{D}})^{-1}[ r_{2}( \overrightarrow{z}
^{\prime }) ,\delta k_{||}( s^{\prime },\overrightarrow{u}
^{\prime }) ] 
\end{eqnarray}

 The result given by eq. 14  due to non-commuting  coordinates   $[r_1(\vec{u}), r_2(\vec{u}') ]_{\mathcal{D}}\neq 0$. We will compare this result with the one  for a    magnetic field  $B$  , $\left[ \delta k_{||}\left( s,\overrightarrow{u}\right) ,\delta k_{||}\left(
s^{\prime },\overrightarrow{u}^{\prime }\right) \right] _{\mathcal{B}}$  given in ref.17. 
 \textbf{ The commutator for the two dimensional densities in the presence of a magnetic field $B(\vec{u})$   perpendicular to the   2DEG has been obtained in ref. [17].}
 
  The modified commutator caused by the magnetic field  (see  eq.  27 in ref. 17)  is:

\begin{eqnarray}
\left[\delta k_{||}\left( s,\vec{u}\right) ,\delta k_{||}\left(
s^{\prime },\vec{u'}\right) \right] _{B} =\left[
\delta k_{||}\left( s,\vec{u}\right) ,\delta
k_{||}\left( s^{\prime },\vec{u'}\right) \right]-\nonumber \\\frac{i e}{h} B(\vec{u})\delta^2 (\vec{u}- \vec{u}')\frac{d}{d \hat{t}(s)}[\delta(K^{0}(s,\vec{u})-K^{0}(s',\vec{u'})]
\label{bac}
\end{eqnarray}

Where $B(\vec{u})$ is the  magnetic field and $\frac{d}{d \hat{t}(s)}=-sin(s)\frac{\partial}{\partial u_{1}} +cos(s)\frac{\partial}{\partial u_{2}} $ is the derivative in the tangential direction  perpendicular to the vector $\widehat{n}(s)$ ( the normal to the Fermi surface) , $\widehat{n}(s)\cdot\nabla=cos(s)\frac{\partial}{\partial u_{1}} +sin(s)\frac{\partial}{\partial u_{2}} $

\textbf{Using the analogy between the vortex field an the external magnetic field  (eq. 15 ) we can represent equation  14 in terms of the parameters given in eq.11.} 

\begin{eqnarray}
\left[\delta k_{||}\left( s,\overrightarrow{u}\right) ,\delta k_{||}\left(
s^{\prime },\vec{u'}\right) \right] _{\mathcal{D}}=\left[
\delta k_{||}\left( s,\vec{u}\right) ,\delta
k_{||}\left( s^{\prime },\vec{u'}\right) \right]-\nonumber \\ \frac{i E(\vec{u},\vec{R})}{D^2}\delta^2 (\vec{u}- \vec{u}')\frac{d}{d \hat{t}(s)}[\delta(K^{0}(s,\vec{u})-K^{0}(s',\vec{u'})]
\end{eqnarray}

\textbf{To conclude the commutator given in equation 16 allows to investigate the physics given in the hamiltonian 12}.

We observe that the Dirac commutator, $\left[ \delta k_{||}\left( s,\overrightarrow{u}\right) ,\delta
k_{||}\left( s^{\prime },\overrightarrow{u}^{\prime }\right) \right]_{\mathcal{D}} \neq 0$
is non-zero for $s\neq s^{\prime }$! 
The Heisenberg equation  of motion will be given by the Dirac bracket.
Due to the fact that different channels $s\neq s^{\prime }$ do not commute,
the application of an external electric field in the $i=s_{\widehat{e}}$
direction will generate a deformation for the F.S. with $s\neq s_{\widehat{e%
}}$, $\delta \dot{k}_{||}\left( s,\overrightarrow{r}^{\prime }\right) =\frac{%
1}{i\hbar }\left[ \delta k_{||}\left( s,\overrightarrow{r}^{\prime }\right)
,H\right] _{\mathcal{D}}$.

Using this formulation we will compute the rectified current.

where $\frac{d}{d \hat{t}(s)}=-sin(s)\frac{\partial}{\partial u_{1}} +cos(s)\frac{\partial}{\partial u_{2}} $ is the derivative according to the tangential direction which is perpendicular to the vector $\widehat{n}(s)$ with, $\widehat{n}(s)\cdot\nabla=cos(s)\frac{\partial}{\partial u_{1}} +sin(s)\frac{\partial}{\partial u_{2}} $

Using this commutation relations given in equation 16  and the hamiltonian given in equation 12 we obtain the equation of motion,

\begin{eqnarray}
\hbar [\frac{d \delta k_{||}( s,\vec{u};t)}{dt}+\frac{\delta k_{||}( s,\vec{u};t)}{\tau}]=V_{F}(s,\vec{u})\left[cos(s)\frac{\partial \delta k_{||}( s,\vec{u};t)}{\partial u_{1}}+sin(s)\frac{\partial \delta k_{||}( s,\cdot\vec{u};t)}{\partial u_{2}}\right]\nonumber\\
-\frac{m\omega^{2}_{0}}{2}\frac{\hbar}{m}\int_{0}^{t}\delta k_{||}( s,\vec{u};t')\,dt + eE_{c}cos(\omega t +\alpha(t))\left[sin(s) +cos(s)\frac{1}{D^{2}}E(\vec{u},\vec{R})\right]
\label{dynamics}
\end{eqnarray}

We have $included$ in the equation of motion a $phenomenological$ $relaxation$ $time$ $\tau$ for the kinetic momentum. 
This equation shows that the direct effect caused by the electric field is proportional to $sin(s)$ with the maximum contribution at the polar angles, $s=\frac{\pi}{2}$ and $s=\pi+\frac{\pi}{2}$ where $V_{F}(s,\vec{u})$ is the Fermi velocity. 
The effect of the $vortex$ is to generates a change in the kinetic momentum perpendicular to the external electric field .
This part is given by the last term. The last term is  restricted to $|\vec{u}-\vec{R}|<D$ and represents the vortex contribution .This term is maximum for the polar angles $s=0$ and $s=\pi$. The maximum effect will be obtained in the region  close to the classical turning point where the Fermi velocity obeys, $V_{F}(s,\vec{u})=\frac{\hbar}{m}{K}_{F}^{0}( s,\vec{u})\approx0$.

The current density in the $i=1$ direction is given by the polar 
integration of $s$  ,$[0-2\pi]$. 
\begin{equation}
J_{1}(\vec{u}) = \frac{e\hbar}{m} \int \frac{\left\vert K^{0}_{F} \left( s,\vec{u} \right) \right\vert }{\left( 2\pi \right)^{2}} \cos(s) \left[ \overrightarrow{K}^{0}_{F}(s,\vec{u}) \delta k_{||}\left( s, \overrightarrow{u}^{\prime } \right) \right]\,ds
\label{curr}
\end{equation}

We introduce the dimensionless parameter $\gamma=\frac{\hbar \omega}{m\omega^{2}_{0} D^{2}}$ which is a function of $\frac{\omega}{\omega_{0}}$ and $D$ the radius of the insulating region. For values of $\gamma<1$ we can solve $iteratively$ the equation of motion and compute the current.

In the equation for the kinetic momentum we have included a phenomenological relaxation time $\tau$. This relaxation time will allow to perform times averages .We only keep single harmonics and neglect higher harmonics of the microwave field.

The iterative solution is given as a series in $\gamma$ and the microwave amplitude $E_{c}$ i.e. $\delta k_{||}( s,\vec{u};t)=\delta k^{(0)}_{||}( s,\vec{u};t)+\gamma\delta k^{(1)}_{||}( s,\vec{u};t)...$.

Solving the equation of motion we determines the evolution of the Fermi surface deformation in the presence of the microwave field.
We substitute the iterative solution $\delta k_{||}( s,\vec{u};t)$   obtained from eq.17 into the current density  formula  given by eq.$18$.

\begin{figure}
% G1.eps: 300dpi, width=1.96cm, height=1.21cm, bb=91 3 322 146
\includegraphics[width=3.5in]{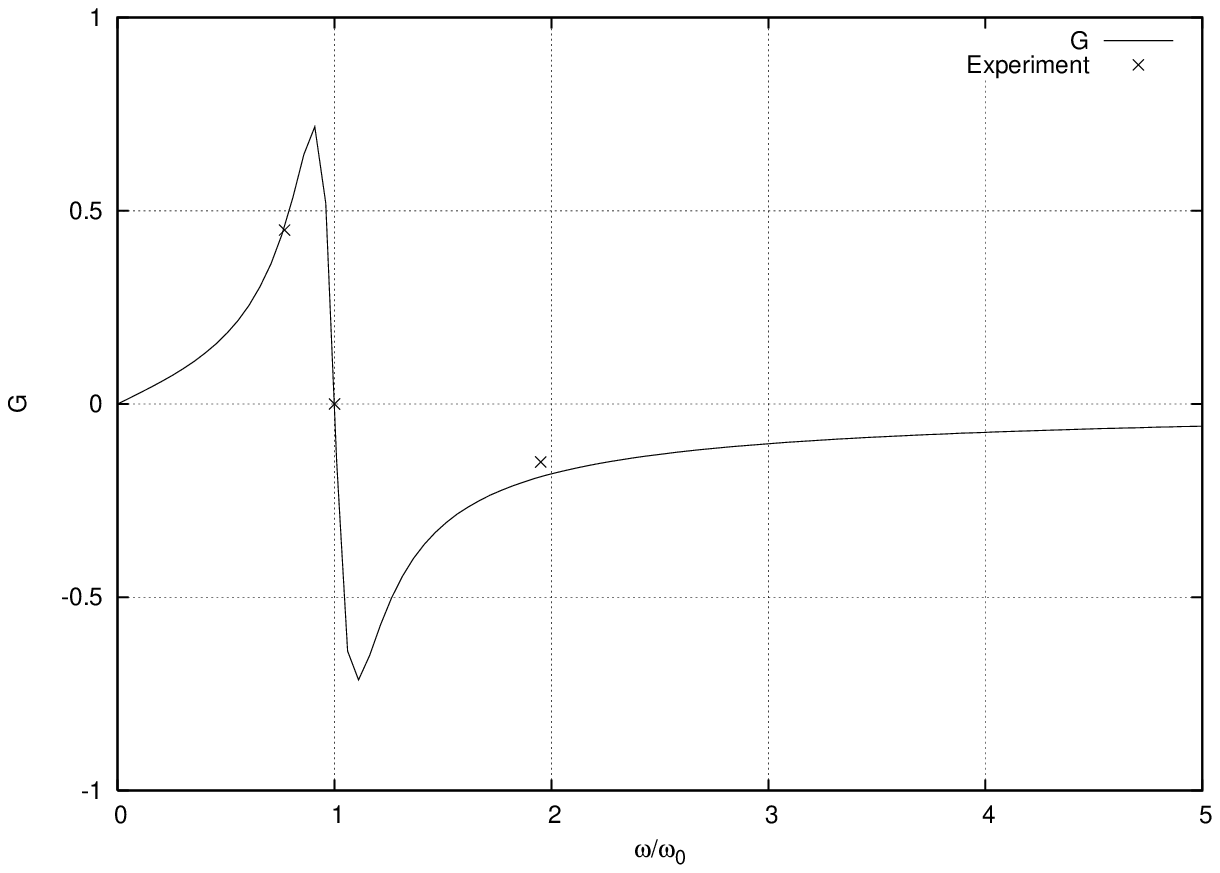}
\caption{}
\label{fig:2}
\end{figure}

 In order to provide a physical interpretation of our theory we will use physical parameters  determined by  the experiment. In the experiment the electronic density is $n \simeq 10^{15} m^{-2}$ this corresponds to a Fermi energy of $E_F \simeq 0.01 eV$, equivalent to a temperature of $T_F \simeq 120 K$ and a Fermi wavelength of $\lambda_F \simeq 0.5 \times 10^{-7} m$. For high mobility GaAs, the typical scattering time is $\tau \simeq 10^{-11} sec$, which corresponds to the mean free path $l=v_{F}\tau $. The ratio between the mean free path and the Fermi wave length obeys the condition, $\frac{l}{\lambda_F} = \frac{v_F \tau}{\lambda_F} = \frac{h \tau}{\lambda_F^2 m} \simeq 30$.
Therefore, we can neglect multiple scattering effects. When the thermal length is comparable with the size of the system $L\simeq \lambda_{thermal} = (\frac{T_F}{T})^{1/2} \lambda_F$, one obtains a ballistic system with negligible multiple scattering.

We describe the confined 2DEG of size $L$ as a system with a parabolic confining potential $V_c(\vec{r}) = \frac{m \omega_0^2}{2} \vec{r}^2$ which has a ``classical turning point'' $L_F$ determined by the condition $\frac{m \omega_0^2 L_F^2}{2} = E_F$. This condition describes the effective physics of a free electron gas of size $L \le L_F$. Demanding that $L_F$ is of the order of the thermal wave length $L_F \simeq \lambda_{thermal}$ determines the confining frequency $\omega_0^2 = \frac{\hbar}{m} (\frac{T}{T_F}) \frac{1}{\lambda_F^2}$. For this condition, we obtain a ballistic regime where $L < L_F \sim 10^{-7} - 10^{-6} m$, $T \sim 1 - 10 K$, $\omega_0 \simeq 10^{10} - 10^{11} Hz$ and $ \tau \simeq 10^{-11} sec$. In order to be able to observe quantum scattering effects caused by the insulating region of radius ``$D$'', we require that the wavelength $\lambda_F$ obeys the condition $D > \lambda_F \simeq 0.5 \times 10^{-7} m$. 

To leading order in $\gamma<1$ and in second order in the microwave amplitude  $E_{c}$ we compute the  rectified D.C. voltage $V_{1, D.C.}$ in the $i=1$ direction. This rectified voltage is defined as $V_{1, D.C.} = I_1 /\sigma$ ($\sigma$ is conductance in the semi classical approximation determined by the transport time which is proportional to the scattering time) .
The current $I_1$ given by
$I_1=\frac{1}{L}\int_{-L}^{L}J_1(\vec{u})\,d^{2}u$ with $L\approx L_{F}$. The microwave field is expressed in terms of an R.M.S. (effective) voltage $V_{R.M.S.} = E_{c} L /\sqrt{2}$ which allows to define a dimensionless voltage in the $i=1$ direction $v_{1, D.C.} = \frac{V_{1, D.C.}}{V_{R.M.S.}} = (\frac{D}{L})^2 \gamma G
(\varphi)$, where $tan(\varphi)=\frac{\omega/\tau}{\omega^{2}-\omega^{2}_{0}}$ with the function $G(\varphi)$ given in figure 2.

For 2DEG, we use typical parameters used in the experiment [7], i.e. electronic density $n\approx 10^{15}m^{-2}$ with a Fermi energy $E_{F}\approx0.01 ev$, $\omega_{0}\approx 10^{10}-10^{11} Hz$, momentum relaxation time $\tau\approx 10^{-11}sec.$ and radius of the insulating region $D>\lambda_{F}\approx 0.5\times10^{-7}m$, with $\gamma\approx 0.7$. We make a single harmonic approximations (neglect terms which oscillate with frequencies $2\omega_{0}$,$3\omega_{0}$ ...) We have used figure $3$ in ref. [7] to extract the voltage changes as a function of the microwave field for a zero magnetic field. Figure $3$ in ref. [7] shows clearly a $change$ of $sign$ when the microwave varies between 1.46 GHz to 34 GHz and vanishes at 17.41 GHz. In figure $2$, we have plotted our results  given by the formula  $ G
(\varphi)$ as a function of the microwave frequency with the rescaled experimental points (see the three points on our theory graph). As shown we find a good agreement of our theory with the experimental results once we choose $\omega_{0}=17.41$ GHz. For frequencies which obeys $\frac{1.46}{17.41} < \frac{\omega}{\omega_{0}} < \frac{34}{17.41}$, we find good agreement with the experimental results. However, for
low frequencies, our theory is inadequate and does not fit the experiment.

\textbf{In conclusion, we can say that the origin of the rectification is the emergence of the non-commuting Cartesian coordinates  and the non-commuting density excitations  are a  result of the vortex field accompanied the  classical turning caused by the  confining potential}. 
Using the modified KacK Moody commutations rule for the density excitations we find that excitations with different polar angles $s$ become coupled.

 Using this  theory we have  explained  the results of the experiment [7] in a region where the magnetic field  was zero.

\end{document}